# Optically switched magnetism in photovoltaic perovskite $CH_3NH_3(Mn:Pb)I_3$

B. Náfrádi[1], P. Szirmai[1], M. Spina[1], H. Lee[2], O. V. Yazyev[2], A. Arakcheeva[1],
D. Chernyshov[3], M. Gibert[4], L. Forró[1] and E. Horváth[1]

[1]Laboratory of Physics of Complex Matter, Ecole Polytechnique Fédérale de Lausanne (EPFL), CH-1015 Lausanne, Switzerland

[2]Institute of Physics, Ecole Polytechnique Fédérale de Lausanne (EPFL), CH-1015 Lausanne, Switzerland

[3]Swiss-Norwegian Beam Lines, European Synchrotron Radiation Facility, 71 Avenue des Martyrs, F-38043 Grenoble Cedex 9, France

[4]DQMP - University of Geneva, 24 Quai Ernest Ansermet, CH - 1211 Geneva 4, Switzerland

**Abstract:** The demand for ever-increasing density of information storage and speed of manipulation boosts an intense search for new magnetic materials and novel ways of controlling the magnetic bit. Here, we report the synthesis of a ferromagnetic photovoltaic $CH_3NH_3(Mn:Pb)I_3$ material in which the photo-excited electrons rapidly melt the local magnetic order through the Ruderman-Kittel-Kasuya-Yosida interactions without heating up the spin system. Our finding offers an alternative, very simple and efficient way of optical spin control, and opens an avenue for applications in low power, light controlling magnetic devices.

The mechanism of magnetic interactions and eventually the magnetic order in insulating and conducting materials are fundamentally different. Diluted localized magnetic (M) ions in insulating materials commonly interact over extended distances by the super-exchange (SE) interaction via atomic orbital bridges through nonmagnetic atoms, *e.g.* oxygen, O. Common schemes for interactions in perovskite structures are the M-O-M, or M-O-O-M-like bridges. The strength and sign (anti- or ferromagnetic, AFM/FM) of these interactions are determined by the geometry of the bonds. Thus, the *in situ* fine-tuning of the interactions is usually difficult because it would call for structural alterations. A limited continuous change is possible by application of pressure[1]. Discrete changes in the lattice are achieved by chemical modifications like replacing the bridging element with halides creating M-Cl-M, M-Br-M or M-I-M bonds[2].

In conducting hosts the long-range double-exchange (DE) or the Ruderman-Kittel-Kasuya-Yosida (RKKY) interactions also come into play between the magnetic M ions. For the RKKY interaction the key control parameters are the density of the localized moments and the density of itinerant electrons. The RKKY coupling strength oscillates between AFM or FM as a function of the M-M distance and of the size of the Fermi surface. These parameters, however, similarly to the case of the SE, are intrinsic to the system and *in situ* modifications are not feasible.

New magnetic materials and efficient, faster ways of controlling the magnetic bit are continuously searched for in order to sustain the needs for ever-increasing density and speed of information storage and manipulation[3-7]. Technologically relevant materials emerge when magnetic interactions of localized and itinerant spins are simultaneously present and compete in determining the ground state. This competition is usually controlled by the carrier concentration and a small external perturbation may result in an extremely large change, for instance, in resistivity. A well-known example is $(La:Sr)MnO_3$ perovskite where at fine-tuned chemical substitutions ferromagnetic DE interactions mediated by chemically doped electrons compete with the antiferromagnetic SE interaction of the parent insulating compound. Consequently a metal-insulator transition and a ferromagnetic order develops [8]. Mechanical and electronic control of the carrier concentration and of the magnetic transition was also shown[1,9,10].

Here we demonstrate an alternative way of controlling the competition of magnetic interactions between itinerant and localized electrons by using visible light illumination. By virtue of photodoping we modulate the carrier concentration and thus the magnetic order in the magnetic photovoltaic perovskite $CH_3NH_3(Mn:Pb)I_3$. This method presents considerable advantages over chemical doping since it is continuously tuneable by light intensity, spatially addressable by moving the illuminating spot and, last but not least, provides a fast switching time (in the ns range required for relaxation of photo-excitations[11,12]). The observed optical melting of magnetism could be of practical importance, for example, in a magnetic thin film of a hard drive, where a small magnetic guide field will trigger a switching of the ferromagnetic moment into the opposite state via the light-induced magnetization melting. This kind of ferromagnetic moment reversal is rapid and represents several indisputable advantages over other optical means of manipulation of the magnetic state reported earlier[3,13-19]. The central ingredient is a high-efficiency photovoltaic material which orders magnetically. Taking advantage of the outstanding light-harvesting characteristics[20] and chemical flexibility[21] of the organometallic perovskite $CH_3NH_3PbI_3$ (hereafter MAPbI$_3$), which has recently triggered a breakthrough in the field of photovoltaics we have developed a magnetic photovoltaic perovskite $CH_3NH_3(Mn:Pb)I_3$ (hereafter MAMn:PbI$_3$), (see Figure 1) by substituting 10% of $Pb^{2+}$ ions with $Mn^{2+}$ ions. This material provides a unique combination of ferromagnetism ($T_C$=25 K) and high efficiency of photoelectron generation. It turns out that in our material these two properties are intimately coupled, thus optical control of magnetism is achieved. Furthermore, we expect this mechanism to be universally present in other magnetic photovoltaics, as well.

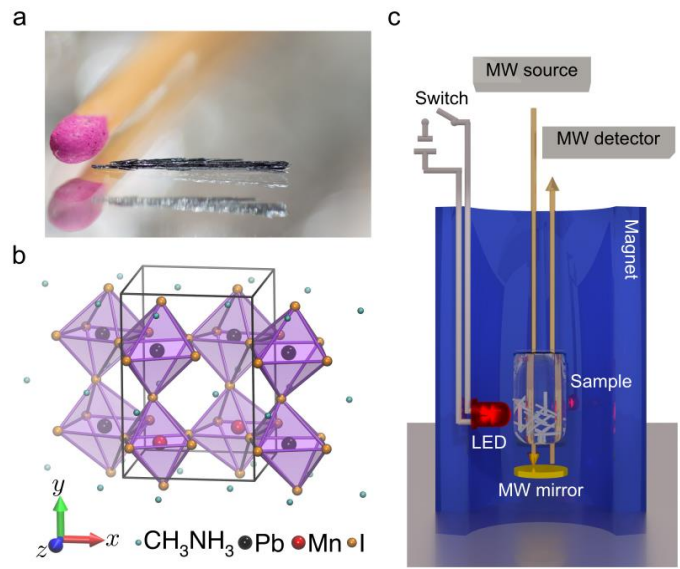

**Figure 1. Sample and measurement configuration.** (a) Photo of a typical $CH_3NH_3(Mn:Pb)I_3$ crystal, 10-15 were assembled for the ESR measurement. (b) Sketch of the crystal structure of $CH_3NH_3(Mn:Pb)I_3$. (c) The experimental configuration for the high-field ESR measurements showing the assembly of small crystals (Sample). The absorption of the microwave field provided by the microwave source (MW source, up to 315 GHz) is monitored (MW detector) in resonant conditions in dark and under illumination in reflection geometry (MW mirror). The light source is a red (λ=655 nm, 4 μW/cm$^2$) Light Emitting Diode (LED) activated by an external switch (Switch).

**Results**

**Magnetic properties in dark**

The substitution of Mn$^{2+}$ ions into the MAPbI$_3$ perovskite network is revealed by synchrotron powder X-ray diffraction and energy dispersive X-ray measurements (Supplementary Figure 1, Supplementary Table 1, and Supplementary Figure 2). Mn$^{2+}$ ions in the host lattice are isoelectronic with Pb$^{2+}$. Hence, they do not dope the system as also confirmed by our first-principles electronic structure calculations discussed below. The substituted sample is semiconducting in dark with few MΩcm resistivity similarly to the parent compound. Moreover, the high level of Mn substitution does not diminish the photocurrent ($I_{ph}$) generation. A strong $I_{ph}$ response is observed below 830 nm wavelength (Supplementary Figure 3). It is worth mentioning that the optical gap decreased relative to the pristine material which facilitates photovoltaic applications. The photocurrent and thus the carrier density can be fine-tuned by the incident light intensity in broad frequency and intensity ranges. The Mn substitution, however, dramatically modifies the magnetic properties of the system as seen by Electron Spin Resonance (ESR) measurements performed in an exceptionally broad 9-315 GHz frequency range (Figure 2 and 3, Supplementary Figure 4-6)[22,23]. The pristine material is nonmagnetic, only ppm level of paramagnetic impurities could be detected. As expected, Mn substitution gives an easily observable signal. At low concentration ESR shows well resolved hyperfine lines indicating the uniform dispersion of Mn$^{2+}$ ions[24] (Supplementary Figure 5). The MAMn:PbI$_3$ sample shows a strong ESR signal (Figure 2 and Supplementary Figure 5) and most importantly, a ferromagnetic order developing below $T_C$=25 K upon cooling in dark. The ferromagnetic order causes a rapid shift of the resonant field, $B_0$, and the broadening of the line width, $\Delta B$, below $T_C$ (Figure 2a and Supplementary Figure 6) which are sensitive measures of the magnetic interactions and the internal magnetic fields[25]. The absence of additional ESR lines in the entire 9-315 GHz frequency range indicates that the magnetic order is homogeneous, the MAMn:PbI$_3$ material is free of secondary phases corroborating with the structural refinement.

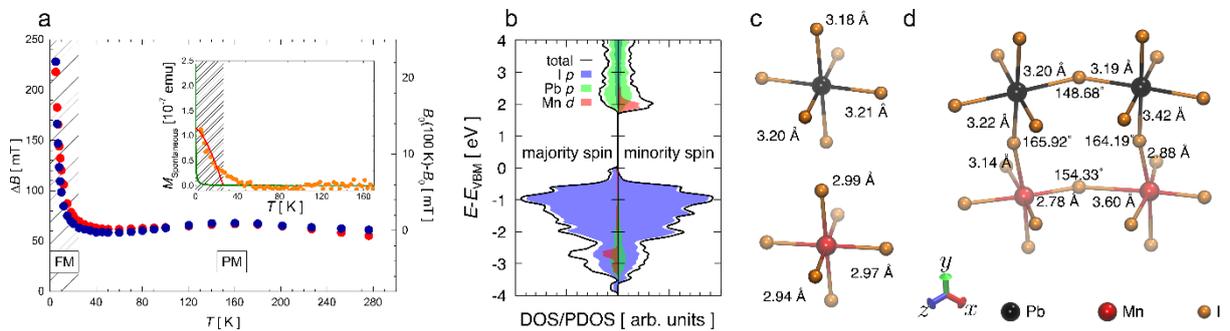

**Figure 2. Magnetic properties of CH$_3$NH$_3$(Mn:Pb)I$_3$ in dark.** (**a**) ESR linewidth (red dots) and resonant field (blue dots, offset by a reference value $B_0$) as a function of temperature recorded at 9.4 GHz. Their temperature independent behavior is characteristic for the paramagnetic phase (PM). The upturn below 25 K corresponds to the on-set of the FM phase. Inset: SQUID magnetometry of MAMn:PbI$_3$. The temperature dependence of the spontaneous magnetization measured in 1.2 μT magnetic field shows a clear increase below $T_C$. The orange line represents the $M_0(1-(T/T_C)^{3/2})$ temperature dependence given by Bloch's Law. (**b**) First-principles calculations of the atomic configurations and magnetic order show total density of states (DOS) and projected density of states (PDOS) calculated for the in-plane model of CH$_3$NH$_3$(Mn:Pb)I$_3$ in its neutral FM configuration. (**c**) Calculated Pb-I and Mn-I distances for a single Mn dopant. (**d**) Calculated bond angles and bond distances for the I mediated superexchange paths in the FM ground state of the in-plane model of CH$_3$NH$_3$(Mn:Pb)I$_3$.

Static magnetization measurements by SQUID (inset to Figure 2a and Supplementary Figure 7) confirm the observations of ESR. The ground state is ferromagnetic as spontaneous magnetization $M_{Spontaneous}$ appears below $T_C$ in the same temperature range where the ESR line shifts and broadens. Moreover, a small magnetic hysteresis appears at low temperatures indicated by the finite remanent magnetization (Supplementary Figure 7). The deviation of $M_{Spontaneous}$ from the mean field description (Bloch's law) in the $T_C$ to $2T_C$ temperature range is characteristic to diluted magnetic systems indicating the phase homogeneity of the system. Temperature and field dependence of $M$ also revealed dominant ferromagnetic correlations at high temperatures, and the presence of magnetocrystalline anisotropy, $K_1=380\times10^4$ J/m$^3$, below $T_C$ (Supplementary Figure 7). The appearance of ferromagnetic order stabilized by short-range SE interactions in the insulating sample at 10% doping levels indicate that both Mn-I-Mn and Mn-I-I-Mn interactions are active to exceed percolation limits.[26]. It should be emphasized that the homogeneous magnetic ordering itself in this insulating photovoltaic perovskite is already a remarkable observation. Such ordering was extensively searched for in homogeneously diluted magnetic semiconductors, and unambiguously observed only in few cases[24,27-29].

This surprising FM order is supported by a rigorous density functional theory (DFT) calculations (see Methods section for calculation details). The model of MAMn:PbI$_3$ was constructed starting from the experimentally determined low-temperature orthorhombic (*Pnma*) crystal structure of undoped material[30], which was then extended to the 2×1×2 supercell. Two Pb atoms in the supercell were replaced by Mn atoms in order to allow investigation of the exchange interactions between Mn dopants. Overall, one Pb atom of eight was substituted, which corresponds closely to the 10% doping concentration of experimentally investigated samples. Three different arrangements of Mn dopants were studied and are shown in Supplementary Figure 8.

The energy differences between the FM and AFM configurations are of the order of 10-20 meV, while the interaction sign varies across the studied models. We found that for the in-plane model (model 2 in Supplementary Figure 8), the FM configuration is the ground state, which is 10.9 meV lower in energy compared to the AFM configuration. The density of states plot calculated for the charge-neutral configuration of in-plane model shows that Mn$^{2+}$ impurities substituting Pb$^{2+}$ ions do not give rise to charge-carrier doping and do not induce any mid-gap states (Figure 2b). The FM interaction is the consequence of the strongly distorted orthorhombic perovskite structure with Mn-I-Mn bond angle significantly reduced to about 150° (Figure 2c-d).

**Melting ferromagnetic order by photoelectrons**

Our major finding is a striking change of the magnetism when the sample is exposed to light illumination at wavelengths lower than the band gap, $\lambda_{edge}=830$ nm (Supplementary Figure 3). Typical ESR absorption spectra taken by light-off and light-on (0.8 μW/cm$^2$) at $T=5$ K are shown in the inset to Figure 3a. The light-on spectrum is considerably narrower and of weaker intensity than the spectrum in dark. The difference between light-on and light-off signals is shown in orange. For the given light intensity, 25 % of the initial spin susceptibility disappears ($\chi_{ESR}$) upon light exposure. In a broad range of illumination intensities, after a threshold value, one can observe a monotonous decrease of the FM part of the signal (Figure 3a). Presumably, below the threshold the photoelectrons fill up some trap sites. At larger intensities, they start to melt the FM state. (The same tendency is observed for $\Delta B$ vs illumination intensity, see Supplementary Figure 9). The change is completely reversible. As $\chi_{ESR}$ is directly proportional to the ferromagnetic volume, the results demonstrate that in a large part of the sample the ferromagnetic order is melted by light illumination. This effect could be closely followed in temperature, as well. The difference between the light-on and light-off signals both in $\Delta B$ and in $\chi_{ESR}$ vary up to $T_C$ (Figure 3b). The narrowing of $\Delta B$ in the remaining magnetic signal, only observed below $T_C$ (Figure 3 and Supplementary Figure 10), is a consequence of the surface melting of the magnetic order, as it is not accompanied by change of $B_0$. The ferromagnetic $\Delta B$ is a strong function of sample shape and size. The light is absorbed in the first few microns of the crystals[11] where the FM is molten so the created magnetic core-shell structure effectively changes the morphology of the sample, thus $\Delta B$. The observations shown in Figure 3b allow us to exclude heating effect by the LED which means that this is an athermal magnetic change induced by photo-excited conduction electrons in the insulating magnetic phase.

**Model the melting of the ferromagnetic order**

On the qualitative basis, one can interpret the light-induced melting of the magnetic structure as the competition between the SE- and the light-induced RKKY-interactions[31]. SE orders the entire sample magnetically in dark. It is known that halide bridges can mediate the interaction between localized $Mn^{2+}$ moments by SE in insulating perovskite crystals[32]. Illumination creates conduction electrons that alter the spin order established by SE as described by the RKKY Hamiltonian[33]. This mechanism is generic to all insulating magnets, where a high efficiency photoelectron generation is present.

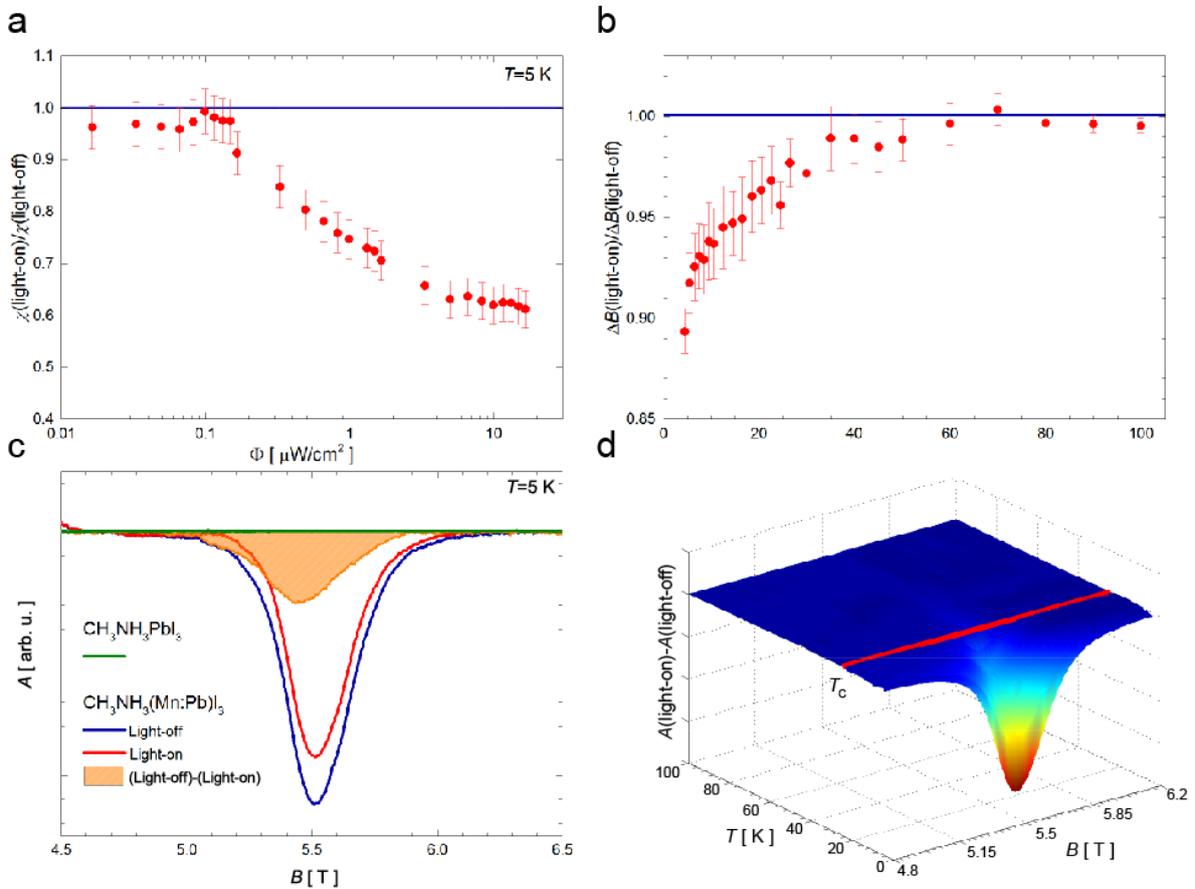

**Figure 3. Illumination effect on the magnetic properties of $CH_3NH_3(Mn:Pb)I_3$ measured by ESR (a)** The intensity change as the function of the illuminating red light intensity $\Phi$ at $T$=5 K. Above a threshold value, the FM part of the signal decreases monotonously. **(b)** Light-on ESR linewidth normalized to the linewidth in dark. The narrowing of the linewidth upon illumination starts below $T_C$. **(c)** ESR spectra at 157 GHz and 5 K of pristine $CH_3NH_3PbI_3$ (green line–no signal), of $CH_3NH_3(Mn:Pb)I_3$ in dark (blue line) coming from the FM phase, and its reduction (red line) upon visible light illumination. The difference between light-off and light-on ESR signal is shown in orange. The effect is accompanied by narrowing of the ESR linewidth upon illumination.

**(d)** Difference of the ESR intensities between the light-off and light-on measurements as a function of temperature. (The third axis shows the resonant field of the signal). The intensity reduction upon illumination is present only below $T_C$=25 K, in the FM phase. Error bars represent the confidence interval of least square fits to the spectra.

Electrical transport measurements in MAMn:PbI$_3$ support this qualitative interpretation. The highly crystalline insulating sample with MΩcm range and thermally-activated resistivity (not shown) transforms even into a metal-like state by the low-intensity red light illumination in a broad temperature range promoting the RKKY interaction. The quadratic magnetoresistance together with the resistivity indicate that even at low temperatures the photoinduced free carrier concentration exceeds $n\sim 2\times 10^{17}$ cm$^{-3}$. Furthermore, the weak, negative magnetoresistance in 0-2.5 T range (above this field a positive, orbital contribution is observed) shows the coupling of conduction electrons to the magnetic moments.

The first idea to model the melting of the FM order by photoelectrons was to consider the competition between the SE- and the RKKY-interactions. This has been performed by DFT calculations (see Methods section for calculation details). Technically, the effect of photoexcited charge carriers was addressed by considering separately electron- and hole-doped models since excitons cannot be described by DFT. Upon doping the in-plane configuration (Supplementary Figure 8) with $2.6\times 10^{20}$ cm$^{-3}$ concentration charge carriers, the ground state changes from FM to AFM with relative energies of 20.4 and 10.9 meV for one hole and for one electron per supercell, respectively. The corresponding total and projected density of states plots for the doped models in their AFM state are shown in Supplementary Figure 11. We have to mention that the carrier concentration used in the modelling is much higher than the measured photoelectron concentration, but the purpose of our calculations is to demonstrate the suppression of the FM order. In fact, calculations with an order of magnitude lower carrier concentrations gave qualitatively identical results.

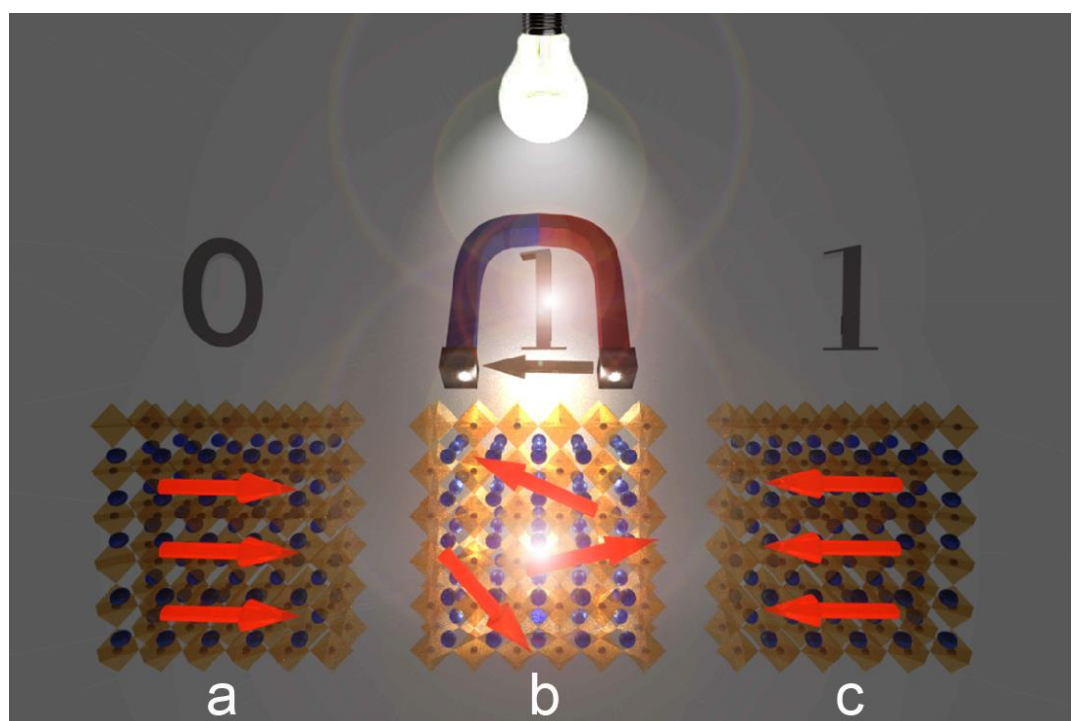

**Figure 4 – Schematic illustration of writing a magnetic bit.** In the dark **(a)** the spin alignment corresponds to a given orientation of the magnetic moment in the FM state, representing a bit. Upon illumination **(b)** the FM order melts and a small magnetic field of the writing head will set the orientation of the magnetic moment once the light is switched off **(c)**.

**Outlook**

As an outlook, the observed optical melting of magnetism could be of practical importance, for example, in a magnetic thin film of a hard drive, where a small magnetic guide field will trigger a switching of the ferromagnetic moment into the opposite state via the light-induced magnetization melting. Its principle is illustrated in Figure 4. This kind of ferromagnetic moment reversal is rapid and represents several indisputable advantages over other optical means of manipulation of the magnetic state reported earlier[3,13-19]. It does not require high-power or femtosecond laser instrumentation, which, besides the complexity of the techniques, raise the stability issue due to photochemistry and fatigue coming from the high intensity and the rapid local thermal cycling of the material[16]. Our method needs only a low power visible light source, providing isothermal switching, and a small magnetic guide-field to overcompensate the stray field of neighboring bits. Although this is a simple and elegant method for magnetic data storage, it has never been discussed in literature, because magnetic photovoltaic materials have not been developed.

**Conclusions**

We have shown the extension of photovoltaics into magnetism by preparing a ferromagnetic MAMn:PbI$_3$. It has been demonstrated that the high-efficiency photocurrent generation by low power visible light illumination results in a melting of the ferromagnetic state and a small local field can set the direction of the magnetic moment. It should be emphasized that this mechanism is radically different from switching the orientation of magnetic domains – here the photoelectrons tune the local interaction between magnetic moments and thus change the magnetic ground state. This study provides the basis for the development of a new generation of magneto-optical data storage devices where the advantages of magnetic storage (long-term stability, high data density, non-volatile operation and re-writability) can be combined by the fast operation of optical addressing. Such a technology should be developed with thin films with higher $T_C$ (which is by far a non-trivial challenge) where the total melting of the magnetism in MAMn:PbI$_3$ could be achieved upon illumination. Last but not least, this study highlights that besides photovoltaics, lasing and LED operation there is one more extraordinary feature of the CH$_3$NH$_3$PbI$_3$ perovskite material.

**Methods**

**Sample preparation:** CH$_3$NH$_3$(Mn:Pb)I$_3$ single crystals were prepared by precipitation from a concentrated aqueous solution of hydriodic acid (57 w% in H$_2$O, 99.99 % Sigma-Aldrich) containing lead (II) acetate trihydrate (99.999 %, Acros Organics), manganese (II) acetate tetrahydrate (99.0 %, Fluka) and a respective amount of CH$_3$NH$_2$ solution (40 w% in H$_2$O, Sigma-Aldrich). The solubility of the Pb- and Mn-acetate provides indirect evidence of the homogeneous distribution of the Mn dopants. A constant 55-42 °C temperature gradient was applied to induce the saturation of the solute at the low temperature part of the solution [21]. Besides the formation of hundreds of submillimeter-sized crystallites (polycrystalline powder) large aggregates of long MAMn:PbI$_3$ needle-like crystals with 5-20 mm length and 0.1 mm diameter were grown after 7 days (Figure 1). Leaving the crystals in open air resulted in a silver-grey to green-yellow colour change. In order to prevent this unwanted reaction with moisture the as synthesized crystals were immediately transferred and kept in a desiccator prior to the measurements. Millimeter size un-doped (CH$_3$NH$_3$PbI$_3$) single crystals were also synthesized and kept as a reference material for qualitative analysis.

**Synchrotron X-ray powder diffraction (XRD)** pattern of the CH$_3$NH$_3$(Mn:PbI)$_3$ sample was measured at room temperature at the Swiss - Norwegian beam lines of the European Synchrotron Radiation Facility (ESRF). The wavelength of the used synchrotron radiation was 0.9538 Å. All data were collected in the Debye–Scherrer geometry with a Dectris Pilatus2M detector. The sample-to-detector distance and the detector parameters were calibrated using a $LaB_6$ NIST reference powder sample. $CH_3NH_3(Mn:PbI)_3$ powders were placed into 10 μm glass capillaries and mounted on a goniometric spinning head. For Rietveld refinement Jana crystallographic program was used. Crystal structure was refined in *I4/mcm* tetragonal space group. Refined atomic parameters of Pb, I, C and N are very similar to those published for $CH_3NH_3PbI_3$ [34]. In addition, H atoms were also localized. The XRD profile together with the results of the Rietveld profile fitting is shown in Supplementary Figure 1. The synchrotron X-ray diffraction profiles revealed a sample without observable secondary phases or phase inhomogeneity.

**Scanning Electron Microscope** images were taken with a MERLIN Zeiss electron microscope. Individual single needle-like crystallites were broken off from the rod like bundles of MAMn:$PbI_3$ for Scanning Electron Microscope micrographs (Supplementary Figure 2). Aluminium pucks were used for sample support. Conducting carbon tape served as electric contact between the sample and the support.

**Energy-dispersive X-ray spectroscopy (EDS).** The elemental composition of the MAMn:$PbI_3$ crystallites were analysed by EDS (accelerating voltage of 8 kV, working distance of 8.5 mm). Samples were mounted on Al pucks with carbon tape with electrical contact to the surface also formed by carbon tape. The measurement was performed with an X-MAX EDS detector mounted at a 35 degrees take-off angle with a SATW window. EDS spectra were obtained at a working distance of 8.5 mm with 8 keV accelerating voltage and a current held at 184 pA. 2048 channels were used for the acquisitions, corresponding to energy of 5 eV per channel. Spectra were acquired over 1573 seconds of live time with detector dead time averaging of 4% and a dwell time per pixel of 500 μs. Quantitative EDS analysis utilized Aztec software provided by Oxford Instrument Ltd.

In order to obtain information on the homogeneity of Mn substitution of the MAMn:$PbI_3$ crystals EDS were performed on several positions on the as-grown surface of the needle-like MAMn:$PbI_3$ crystallites. For the purpose of gathering bulk information as well EDS spectrum were taken also on broken-off surfaces. These experiments systematically yield $(Mn_{0.1}Pb_{0.9})I_3$ stoichiometry indicating homogeneous Mn substitution.

**Electron spin resonance spectroscopy (ESR).** Polycrystalline assembly of 10-15 rod like MAMn:$PbI_3$ samples with typical 1 mm×0.1 mm×0.1 mm are sealed in a quartz capillary. ESR at 9.4 GHz microwave frequency was performed on a Bruker X-band spectrometer. A conventional field modulation technique was employed with lock-in detection which results the first derivative of the ESR absorption spectra. Experiments in the mm-wave frequency range were performed on a home-built quasi-optical spectrometer operated at 75, 105, 157, 210, and 315 GHz frequencies in 0-16 T field range (Figure 1). The spectral resolution of ESR is linearly proportional to the frequency, thus we extended the precision of ESR by about a factor 30 compared to X-band ESR technique. More details about the setup can be found in [22,23]. A red LED was placed underneath the sample as a light source. Magnetic field strength at the sample position was calibrated against a $KC_{60}$ standard sample. In contrast to the low-field ESR experiments, at millimeter-wave frequencies a microwave power chopping was combined with lock-in detection. This detection scheme results directly the ESR absorption signal instead of its first derivative. The working principles of the two methods are shown in Supplementary Figure 4.

Supplementary Figure 5 (a-e) compares pristine $MAPbI_3$ with 1% and 10% substituted MAMn:$PbI_3$ at room temperature. Pristine $MAPbI_3$ crystals show no intrinsic ESR signal. Only low, ppm levels of paramagnetic impurity centers were observed (Figure 3 and Supplementary Figure 5). In contrast, Mn substitution to MAMn:$PbI_3$ results in a strong ESR signal.

The presented ESR experiments prove that the magnetic transition is not driven by temperature change. ESR unambiguously demonstrates that we are not dealing with a temperature effect. By ESR at each temperature one obtains the spin susceptibility, ESR linewidth and resonance field *simultaneously*. All 3 parameters are strongly temperature-dependent as shown in Figure 2 and Supplementary Figure 6. Temperature change modifies all the three parameters concurrently. Our careful ESR experiments performed in dark provide us an internal thermometer. Increasing the temperature by ~1 K would change all 3 aforementioned ESR parameters *simultaneously*. Switching on the light does not show this effect as demonstrated in Figure 3. It changes the spin susceptibility, ESR linewidth and resonance field by an amount that corresponds to different temperature changes. This cannot be explained by a temperature effect. This shows unambiguously that we are not dealing with a temperature effect. Accordingly, the phenomenon we discovered is an athermal effect. Instead, we would like to point out that our ESR experiments are demonstrating the change of Curie temperature with photo-excitation. The spin-susceptibility, measured by the ESR intensity, at $T<T_C$ decreases by about 25% upon light illumination (see Figure 3b and 3d). This demonstrates the disappearance of 25% of the FM volume upon illumination. It means that in that 25% volume the $T_C$ decreased from 25 K to below 5 K the lowest temperature in our experiments.

The spectra at 1% $Mn^{2+}$ concentration consist of two signals. One set of sextet lines and an about 50 mT broad line (see Supplementary Figure 5). The sextet signal is characteristic of a hyperfine splitting of $^{55}Mn$ with $g = 2.001(1)$ *g*-factor and $A_{iso} = 9.1$ mT hyperfine coupling constant [24]. This spectrum corresponds to both allowed (sextet) and forbidden (broad component) hyperfine transitions between the Zeeman sublevels. It is characteristic to $Mn^{2+}$ ions in octahedral crystal fields. Since strong forbidden transitions are observed, $Mn^{2+}$ ions do not occupy strictly cubic sites, as strictly cubic centers have zero probability of forbidden transitions, rather distorted octahedral sites. The well resolved hyperfine also testifies the homogeneous distribution of Mn ions in $MAMn:PbI_3$.[24]

These ESR characteristics are in good agreement with both powder X-ray diffraction and DFT calculations showing distorted octahedral Mn coordination. The ESR spectra of $MAMn:PbI_3$ at high $Mn^{2+}$ concentration (10%) consist of one broad ESR line only. This is a common resonance of both allowed and forbidden transitions. We explain the uniformity of the g-factor by strong exchange narrowed spin-orbit interaction dominated line width of the $Mn^{2+}$ ions. Following the calculations of ref. [35] and assuming a spin orbit width contribution of the order of $(\Delta g/g)J$, yields a value of the order of 100 K for exchange integral *J*.

The broad ESR and isotropic *g*-factor is strongly intrinsic for the system. We find no evidence of frequency dependence at high temperatures in the 9-315 GHz frequency range. The field and temperature independent $\Delta B$ and $B_0$ is characteristic to exchange coupled paramagnetic insulators. Below 25 K, both $\Delta B$ and $B_0$ acquires strong temperature dependence indicative of ferromagnetic ordering. The shift in $B_0$ measures the temperature dependence of the internal ferromagnetic field of $MAMn:PbI_3$. $\Delta B$ scales to $B_0$ at all measure fields and temperatures (see Figure 2 and Supplementary Figure 6) indicating an inhomogeneous broadening induced by spatial distribution of the local internal ferromagnetic field. The inhomogeneity of the local internal ferromagnetic field is partially of geometrical origin. The demagnetizing field of our irregularly shaped particles is inhomogeneous. Additionally, the statistical fluctuations of the Mn concentration across the sample also increase the inhomogeneity by modulating the strength of the ferromagnetic order.

The magnetic phase purity can be further confirmed by comparing the two described ESR method. The microwave chopping method (see Supplementary Figure 4b), which yields the integrated ESR signal, would reveal possible broad ESR signals. However, Figure 3b and Figure 3d show the absence of any broad magnetic impurity signals. The magnetic field modulation method (see Supplementary Figure 4a), would help to identify narrow signals with a linewidth in the order of the modulation. Supplementary Figure 5f proves the absence of the narrow impurity signals as well.

We note here that the signal of itinerant electrons generated by the illumination is not detected, either. Detection of a so-called conduction electron spin resonance (CESR) line would be a major challenge (see, e.g., ref [24,36]). The two main difficulties in the order of the importance are: (i) low Pauli spin susceptibility of a CESR signal, and (ii) the spin-orbit coupling provokes a broadening in the signal[37,38].

In our system, the presence of conduction electrons can be excluded in dark, as MAMn:PbI$_3$ is an insulator without light. We would only observe the CESR signal in the presence of photoexcited carriers upon illumination. As seen in Figure 3b and in Supplementary Figure 5f, however, we do not observe the CESR upon illumination. In our case, both issues of CESR detection are significant. The weak illumination results in the small spin-susceptibility of the generated conduction electrons. In fact, the expected spin susceptibility of the CESR (Pauli susceptibility) of the photoexcited state is 5-6 orders of magnitude smaller than the paramagnetic Mn$^{2+}$ ESR signal. Furthermore, the large spin-orbit coupling broadens this small signal. These two effects prevent the observation of the CESR.

Furthermore, the precursor Mn-acetate used for the Mn substitution has markedly different ESR spectra from the substituted material, thus inclusions of Mn-acetate islands can be excluded, as well.

**SQUID magnetometry** experiments reveal that the temperature dependence of spontaneous magnetization, the defining macroscopic property of ferromagnetism, appears below 50 K and dramatically enhances below 25 K (Figure 2). The theoretical behaviour of the paramagnetic magnetization in the same conditions is shown by the blue dashed line. Clearly, the spontaneous magnetization is orders of magnitude greater compared to a paramagnetic magnetization, testifying the ferromagnetic order.

The mean field theory of spontaneous magnetization is described by Bloch's law, which states that $M_{spontaneous} \sim 1-(T/T_C)^\alpha$ with $\alpha=3/2$ (orange line in Figure 2). However, deviations from the mean-field exponent are recurrent, e.g., Iron and Nickel show critical exponents $\alpha$ of 0.34 and 0.51, respectively. Similarly, the spontaneous magnetization in MAMn:PbI$_3$ deviate from the mean field value (see Figure 2). The deviation from Bloch's Law is indicative of the presence of strong magnetocrystalline anisotropy. The primary source of magnetocrystalline anisotropy is the spin-orbit interaction, which is strong due to the involvement of Pb and I atoms.

In Supplementary Figure 7a, we show the temperature dependence of the magnetization cooled in 1 T external field. In agreement with the appearance of the remanent magnetization in the zero field-cooled experiments, we find a Curie-Weiss temperature of $T_{CW}$=14 K characteristic to predominant ferromagnetic correlations. At low temperatures, however, the magnetization is suppressed relative to the isotropic Curie-Weiss behaviour. This is characteristic to the presence of magnetocrystalline anisotropies with perfect agreement with the observed deviation of the spontaneous magnetization from the mean-field description.

The magnetic field dependence of the magnetization measured at $T$=2 K up to 7 T magnetic field (see Supplementary Figure 7b) shows a steady increase of magnetization with about $H_S=2K_1/M_S$ = 9 T saturating magnetic field. This again underlines the presence of magnetic anisotropy of $K_1$=380×10$^4$ J/m$^3$ at $T$=2 K. Note that this value is in the same range as those found at room temperature in hematite ($K_1$=120×10$^4$ J/m$^3$) and for YCo$_5$ ($K_1$=550×10$^4$ J/m$^3$).

Finally, the temperature dependence of the remanent magnetization measured by decreasing the magnetic field from 7 T shows small value in agreement with the magnetization isotherms. The temperature behaviour is similar to the behaviour of the spontaneous magnetization, and it increases below 50 K (inset of Figure 2).

These SQUID experiments undoubtedly reveal the existence of magnetic order of our MAMn:PbI$_3$ compound. It also shows high magnetic phase purity. No sign of additional magnetic or nonmagnetic phase was detected in perfect agreement with the multi-frequency ESR investigations.

**Photocurrent spectroscopy.** For photocurrent spectra a low intensity monochromatic light was selected by a MicroHR grid monochromator from a halogen lamp. The wavelength resolution (FWFM) of the 600 gr/mm grating was 10 nm. The photo excited current was measured by a two-terminal method at fixed bias voltage of 1 V while the wavelength was stepwise changed (Supplementary Figure 3). Measurements were performed on pristine MAPbI$_3$ and Mn doped MAMn:PbI$_3$. The band gap energy was determined by fitting a Fermi-Dirac distribution to the data. The resulting gap energies at room temperature are 783±1 nm and 829±1.4 nm for the MAPbI$_3$ and MAMn:PbI$_3$, respectively. The intrinsic width of the Fermi-Dirac distribution for both systems is thermally broadened. This indicates that the Mn substitution is homogeneous. Mn clustering would cause broadening of the band edge. It is also worth mentioning the strong, about 46 nm upshift of the band edge upon Mn substitution since the gap of MAMn:PbI$_3$ is reduced relative to MAPbI$_3$. Mn substitution presents an alternative route to extend the light absorption range, hence increase photocell efficiencies. The temperature dependence of the photocarrier generation in 50-300 K temperature range was also studied in a closed-cycle cryostat equipped with an optical window (Supplementary Figure 3). The gap energy increases by decreasing temperature due to thermal expansion, however, the photocarrier generation of MAMn:PbI$_3$ remains effective down to the lowest studied temperatures.

**First-principles electronic structure calculations.** To corroborate the experimental findings, we carried out first-principles electronic structure calculations in the framework of density functional theory [39,40] as implemented in the Quantum ESPRESSO package [41]. The exchange-correlation energy is given by the Perdew-Burke-Ernzerhof generalized gradient approximation [42] while the electron-ion interactions are treated by using the ultrasoft pseudopotentials [43] which have been previously published[44]. Wave functions and charge densities are expanded using the plane-wave basis sets with kinetic energy cutoffs of 40 Ry and 320 Ry, respectively. The Brillouin zone (BZ) is sampled using 3×4×3 Monkhorst-Pack meshes of special **k**-points [45]. The plane-wave cutoffs and **k**-point meshes are chosen to ensure the convergence of total energies within 10 meV. When performing calculations on charged models, a compensating jellium background was introduced in order to avoid the spurious divergence of electrostatic energy [46].

The models of Mn-doped CH$_3$NH$_3$PbI$_3$ were constructed starting from the experimentally determined crystal structure of undoped material (orthorhombic phase, space group *Pnma*) [30], which was then extended to the 2×1×2 supercell by doubling the lattice constants along the *a* and *c* directions. Two Pb atoms in the supercell were replaced by Mn atoms in order to allow investigating the exchange interactions between Mn dopants. Overall, one Pb atom of eight was substituted, which corresponds closely to the doping concentration of experimentally investigated samples (10 %). Three different arrangements of Mn dopants, referred to as top, in-plane, and diagonal, are shown in Supplementary Figure 8. Atomic coordinates of all these three configurations were optimized to the residual ionic forces smaller than 0.02 eV·Å$^{-1}$, whereas the lattice parameters were kept fixed. For each configuration both the ferromagnetic (FM) and antiferromagnetic (AFM) arrangements of local magnetic moments of Mn atoms were investigated. Our calculations show that optimization of the internal atomic coordinates is crucial for reproducing the relative energies of FM and AFM configurations. Indeed, substitution of Mn atoms for Pb atoms leads to a pronounced lattice distortion around the Mn dopants due to different ionic sizes of $Mn^{2+}$ and $Pb^{2+}$. Specifically, the Mn-I distances are about 2.9 Å, whereas the Pb-I distances are about 3.2 Å (Figure 2b-c).

For all considered arrangements of Mn dopants, the energy differences between the FM and AFM configurations are of the order of 10-20 meV. We found that for model 2 (in-plane, Supplementary Figure 8), the FM configuration is the ground state, which is 10.9 meV lower in energy compared to the AFM configuration. Due to intrinsic limitations of density-functional-theory calculations, the effect of photoexcited charge carriers was addressed by considering separately electron- and hole-doped models. One has to emphasize that the DFT calculations correspond to a 0 K case and fixed number of photoelectrons. At finite temperatures and variable carrier density between the FM and AFM configurations it is reasonable to expect a paramagnetic state as seen in the experiment.

**DC resistivity and magnetotransport under illumination** were performed with the same light conditions as the ESR experiments. Resistivity and magnetoresistance were measured in a standard 4-terminal configuration in the 5-300 K temperature and 0-16 T magnetic field range. In dark, the resistivity of the samples is in the MΩcm range and show thermally-activated character (not shown). Under red light illumination, the resistivity monotonically drops by lowering temperature. At the structural transition temperature around 150 K, however, the resistivity discontinuously jumps. Magnetoresistance at low temperatures increases quadratically by increasing magnetic field. In the carrier/exciton ratio study of D'Innocenzo et al[47], it was suggested that free charge carriers are predominant in perovskite solar cells at room temperature. Hence, the perovskites possess large built-in fields which can effectively drift photogenerated carriers to avoid charge recombination. These are in perfect agreement with our magnetoresistance and photocurrent spectroscopy measurements. The resistivity decreases by a factor more than 400 by cooling from 300 K to 30 K. This clearly indicates the presence of free carriers at temperatures relevant to FM melting. The photocurrent generation is an active process. Also, the photocurrent spectra at 50 K clearly shows that free carriers are readily excited in our experiments.

Our experiments were performed under continuous illumination, which implies a constant number (time independent after a few fs transient) of out-of-equilibrium photoexcited carriers next to the thermalized free carriers and excitons.

**Data availability**. The data that support the findings of this study are available from the corresponding authors upon request.

**Supplementary Information** accompanies this paper.

**Acknowledgements**

We are grateful to J.M. Triscone for facilitating SQUID experiments. This work was partially supported by the Swiss National Science Foundation (Grant No. 200021_144419) and ERC Advanced Grant "PICOPROP" (Grant No. 670918). H.L. and O.V.Y. thank the Swiss NSF (Grant No. PP00P2_133552), the ERC starting grant "TopoMat" (Grant No. 306504) and NCCR-MARVEL. First-principles computations have been performed at the Swiss National Supercomputing Centre under Project No. s515.

**Author Information,**

Laboratory of Physics of Complex Matter, Ecole Polytechnique Fédérale de Lausanne (EPFL), CH-1015 Lausanne, Switzerland
Bálint Náfrádi, Péter Szirmai, Massimo Spina, Alla Arakcheeva, László Forró, Endre Horváth
Institute of Physics, Ecole Polytechnique Fédérale de Lausanne, CH-1015 Lausanne, Switzerland
Hyungjun Lee, Oleg V. Yazyev
Swiss-Norwegian Beam Lines, European Synchrotron Radiation Facility, 71 Avenue des Martyrs, 38043 Grenoble Cedex 9, France
Dmitry Chernyshov
DQMP - University of Geneva, 24 Quai Ernest Ansermet, CH - 1211 Geneva 4, Switzerland
Marta Gibert

Reprints and permissions information is available at www.nature.com/reprints. The authors declare no competing financial interests.
Correspondence and requests for materials should be addressed to nafradi@yahoo.com or to laszlo.forro@epfl.ch

**Author Contributions**

L.F. initiated the research. B.N. suggested the idea, coordinated the efforts and designed the experiment. E.H. synthesized the substance. P.Sz., M.S., D.C., M.G. and B.N. derived the experimental results. A.A. analysed the XRD pattern. H.L. and O.V.Y. performed first principle calculations. All authors discussed the results and contributed to the writing of the manuscript.

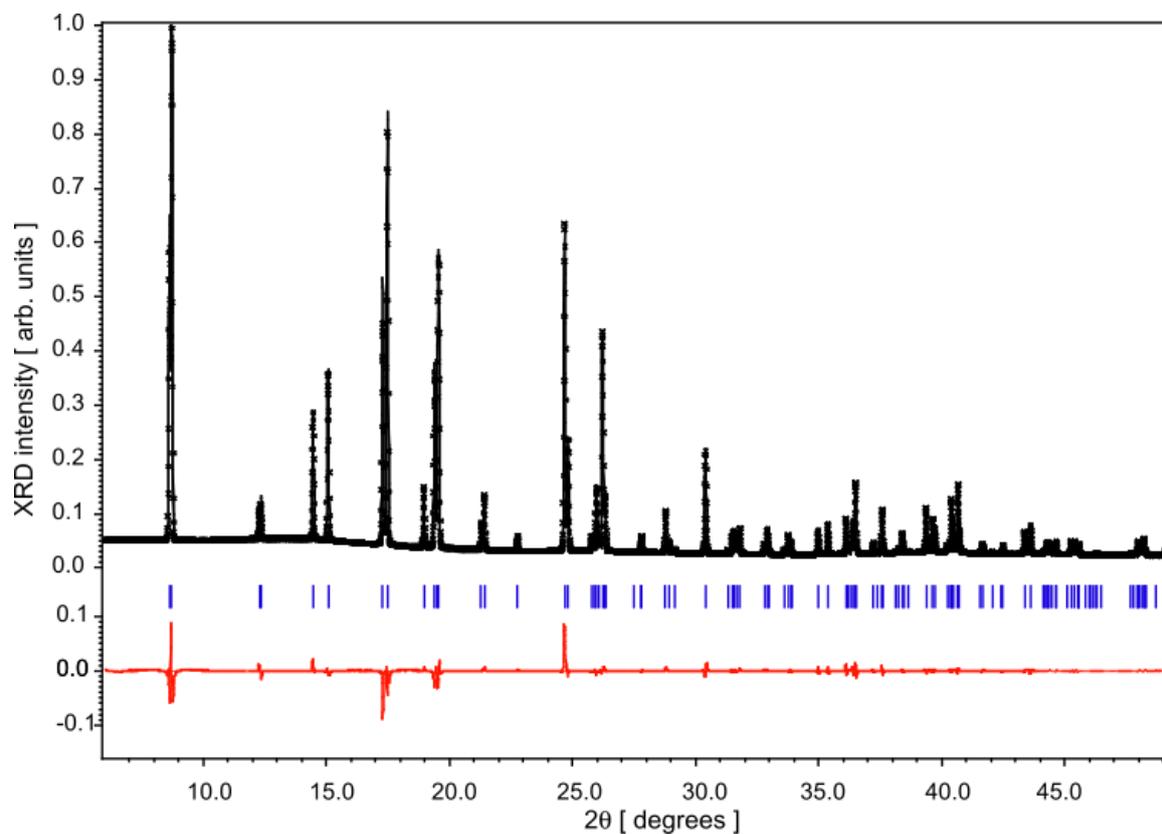

**Supplementary Figure 1 - Synchrotron powder X-ray diffraction.** Room temperature synchrotron powder X-ray profile of MAMn:PbI$_3$ (wavelength of the synchrotron radiation is equal to 0.9538 Å). Stars and solid and thin lines (black) correspond to experimental data and calculation, respectively. Deviation from the fit is shown in red. Strips (blue) indicate positions of the Bragg reflections. The Rietveld refinement shows a perfectly single phased material: MAMn:PbI$_3$ sample is free of PbI$_2$, Mn clusters or any other impurity.

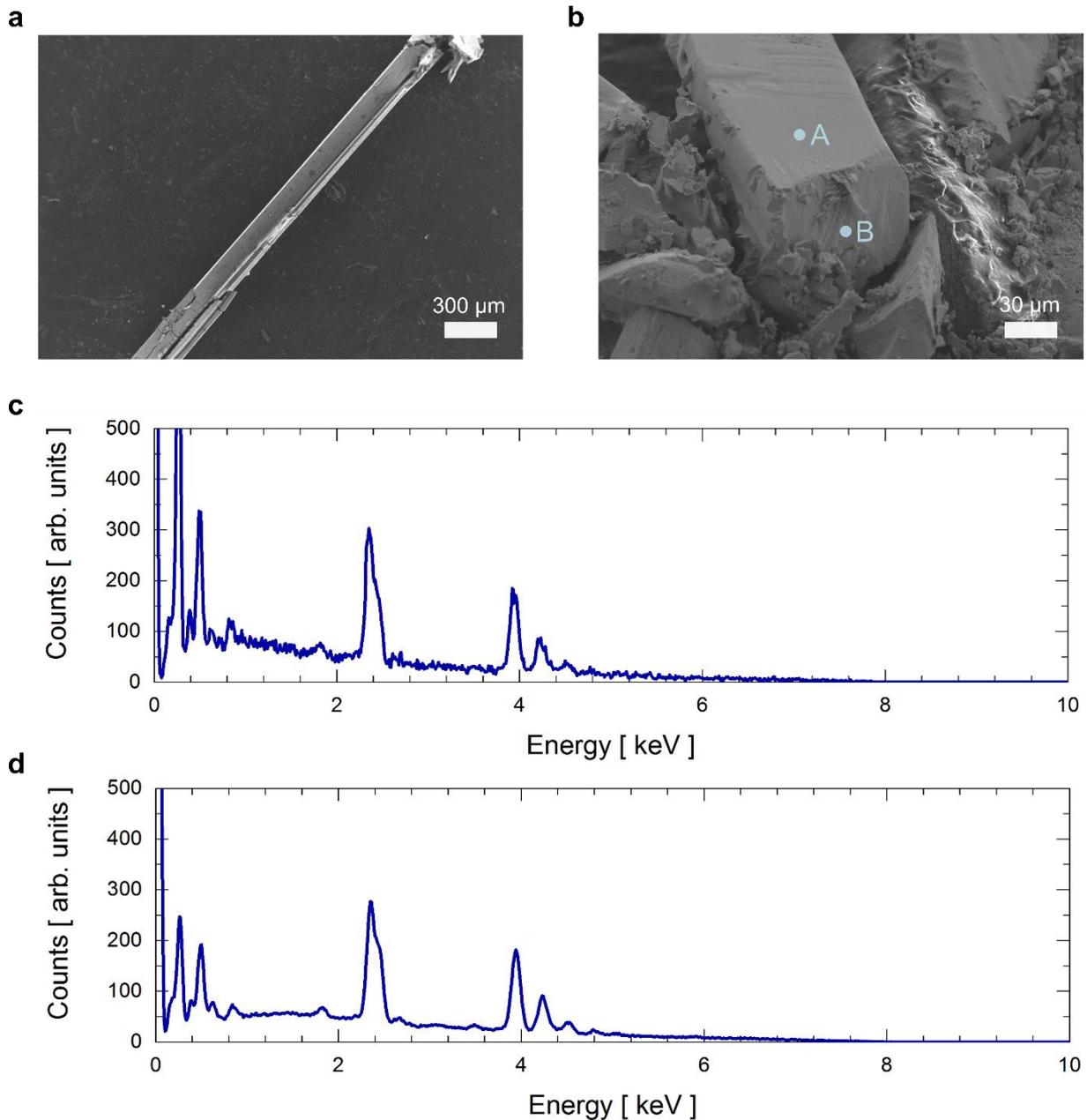

**Supplementary Figure 2 - Energy dispersive X-ray spectroscopy.** (**a**) SEM micrograph of a typical MAMn:PbI$_3$ single crystal of several mm in length and 100×100 µm$^2$ in cross-section. (**b**) Zoom on a broken section of the needle shown in **a**. A and B are the positions where the EDS spectra were obtained. (**c-d**) EDS sum spectra obtained at the as-grown and broken surfaces indicated by A (**c**) and B (**d**), respectively in **b**. The stoichiometry at both regions is Pb$_{0.9}$Mn$_{0.1}$I$_3$, testifying the homogeneous bulk substitution of Mn ions.

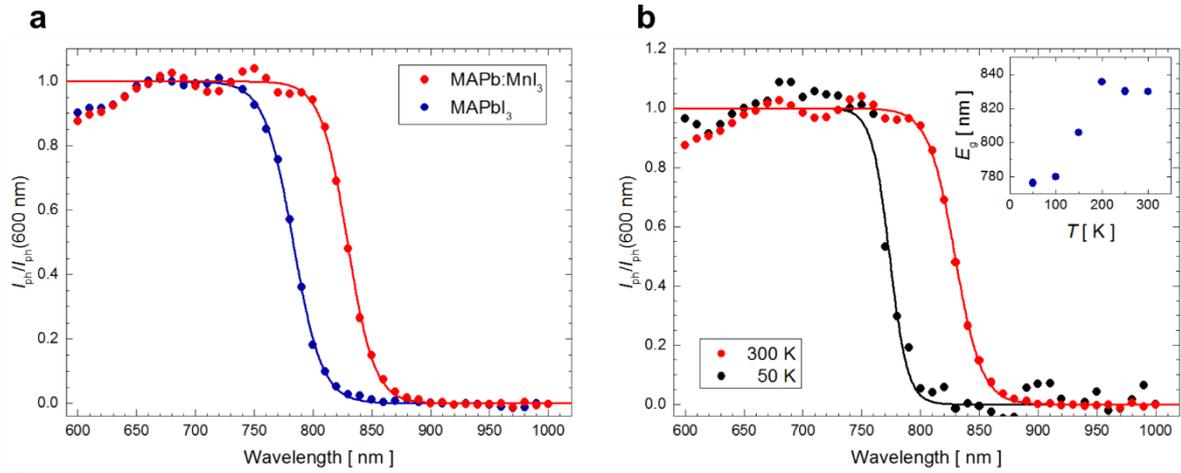

**Supplementary Figure 3 - Photocurrent spectra.** (a) Photocurrent of MAMn:PbI$_3$ (red symbols) and MAPbI$_3$ (blue symbols) at fixed bias voltage of 1 V measured as a function of photon energy at 300 K. The strong photocurrent generation above the optical band gap of ~830 nm of MAMn:PbI$_3$ is red shifted by about 46 nm relative to that of the pristine MAPbI$_3$ material (783 nm). Lines are fits to modelling the band edge by the Fermi-Dirac distribution and its thermal broadening. (b) Comparison of the $T$=50 K (black) and $T$=300 K (red) photocurrent spectra of MAMn:PbI$_3$. Inset shows the temperature evolution of the bandgap ($E_g$) obtained from photocurrent spectroscopy.

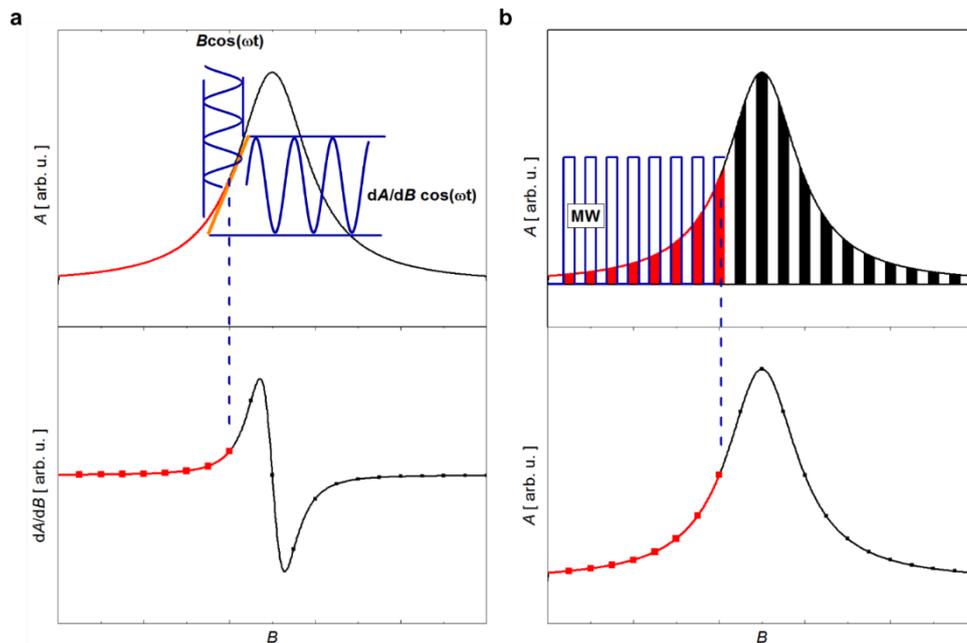

**Supplementary Figure 4 - Basic principle of ESR signal detection.** (**a**) Conventional magnetic field modulation used in 9.4 GHz ESR experiments. Upper curve represents the ESR absorption *A* as a function of magnetic field *B*. The modulation magnetic field $B\times\cos(\omega t)$ and the resulting modulated microwave absorption power $dA/dB\times\cos(\omega t)$ are also illustrated. Lower panel depicts the *first derivative* d*A*/d*B* signal of the ESR absorption line *A* after lock-in detection. (**b**) Microwave (MW) chopping detection used for 105 and 157 GHz ESR experiments. The microwave radiation is periodically switched on/off (blue line). Accordingly, the ESR absorption signal is modulated as shown by the red shaded area. The lower panel presents the resulting absorption ESR line *A* after lock-in detection.

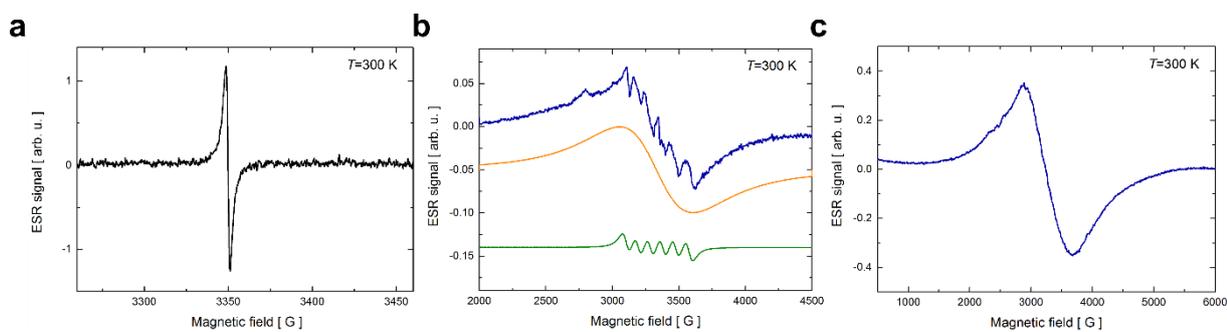

**Supplementary Figure 5 – Room-temperature 9.4 GHz ESR spectra.** **(a)** Spectrum of pristine MAPbI$_3$. Only a weak paramagnetic impurity signal is observed characteristic of ppm level defect concentration. **(b)** Spectra of MAMn:PbI$_3$ with low (~1%) Mn concentration. A forbidden hyperfine signal (orange) and allowed hyperfine sextet line (green) of the Mn$^{2+}$ reproduce the observed signal well (blue). The well-resolved hyperfine structure indicates the homogeneous dispersion of the Mn$^{2+}$ ions. **(c)** Spectrum of MAMn:PbI$_3$ with high (10%) Mn concentration.

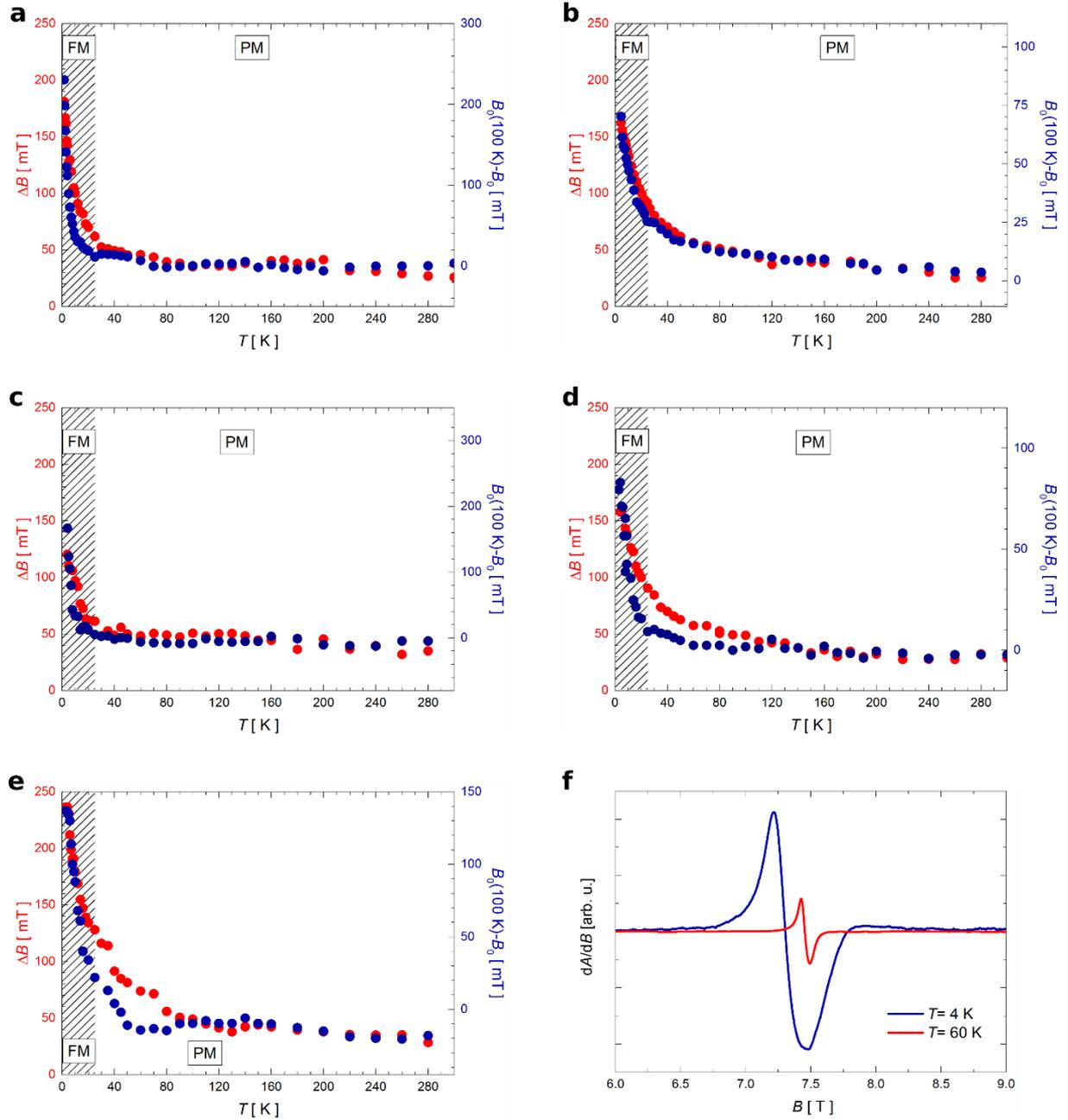

**Supplementary Figure 6** - **Multifrequency ESR properties of MAMn:PbI$_3$.** (**a-e**) ESR at 75 (**a**), 105 (**b**), 157 (**c**), 210 (**d**), and 315 GHz (**e**) frequencies were measured as a function of temperature. The temperature dependence of the linewidth (red) scales with the temperature dependence of the ESR shift $B_0(100\ K)-B_0$ (blue) showing that both quantities measure the local dipole field distribution of the polycrystalline ferromagnetic material. FM and PM show the ferromagnetic (shaded area) and paramagnetic state, respectively. (**f**) Comparison of the first-derivative ESR spectra measured below (*T*=4 K) and above $T_C$ (*T*=60 K). Absence of narrow ESR components below $T_C$ proves the high magnetic phase purity.

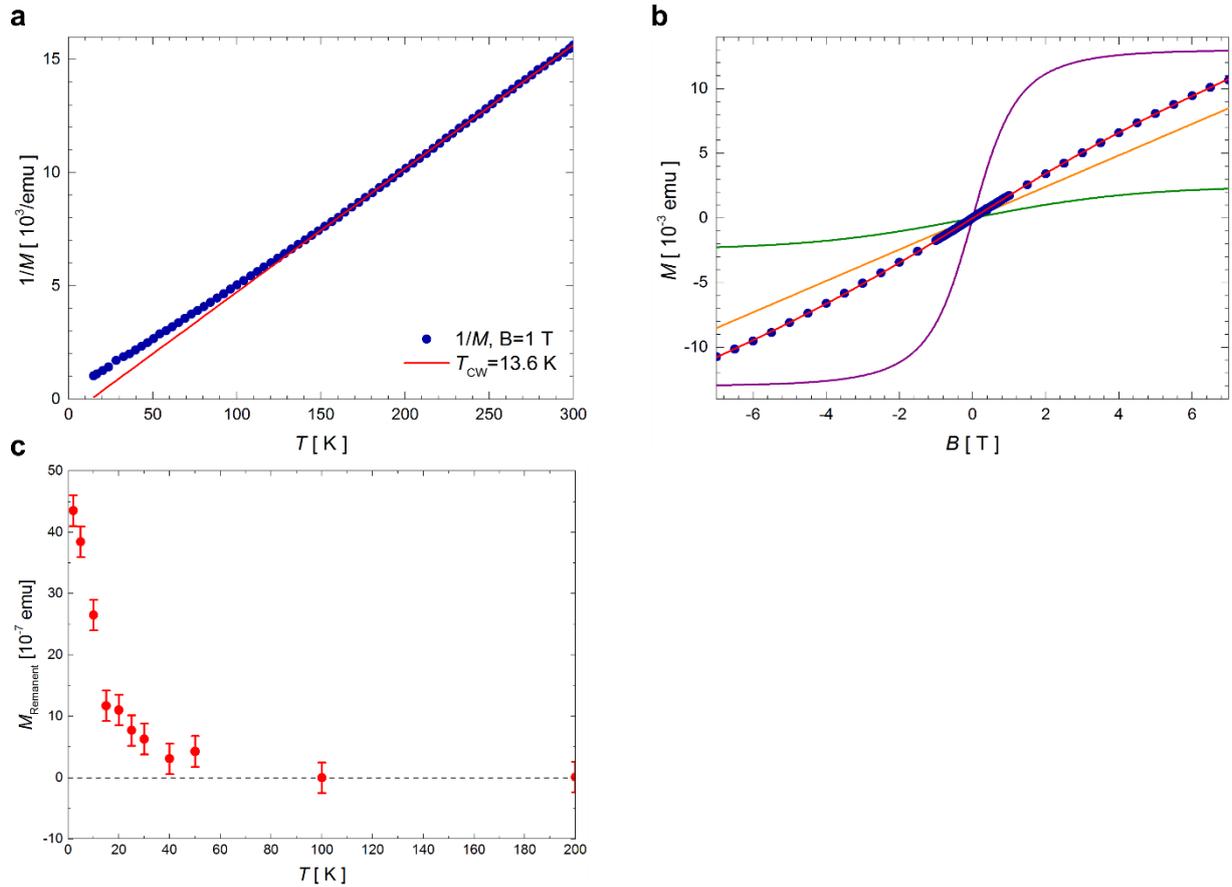

**Supplementary Figure 7- SQUID magnetometry of MAMn:PbI$_3$**. **(a)** Temperature dependence of 1/$M$ cooled in 1 T magnetic field. Line presents the Curie-Weiss fit which reveals predominant ferromagnetic correlations with a Curie-Weiss temperature of $T_{CW}$=13.6 K. **(b)** Magnetization measured a $T$=2 K after a field cooled process in 7 T. The field dependence is remarkably well described by (red line) a ferromagnetic powder with $K_1$=380×10$^4$ J/m$^3$ (orange) and with a small magnetic domain contribution (green). The observed behaviour is clearly distinct from a paramagnet case (purple). **(c)** Remanent magnetization as a function of temperature. Error bars represent the confidence interval of least square fits to the $M(H)$ curves.

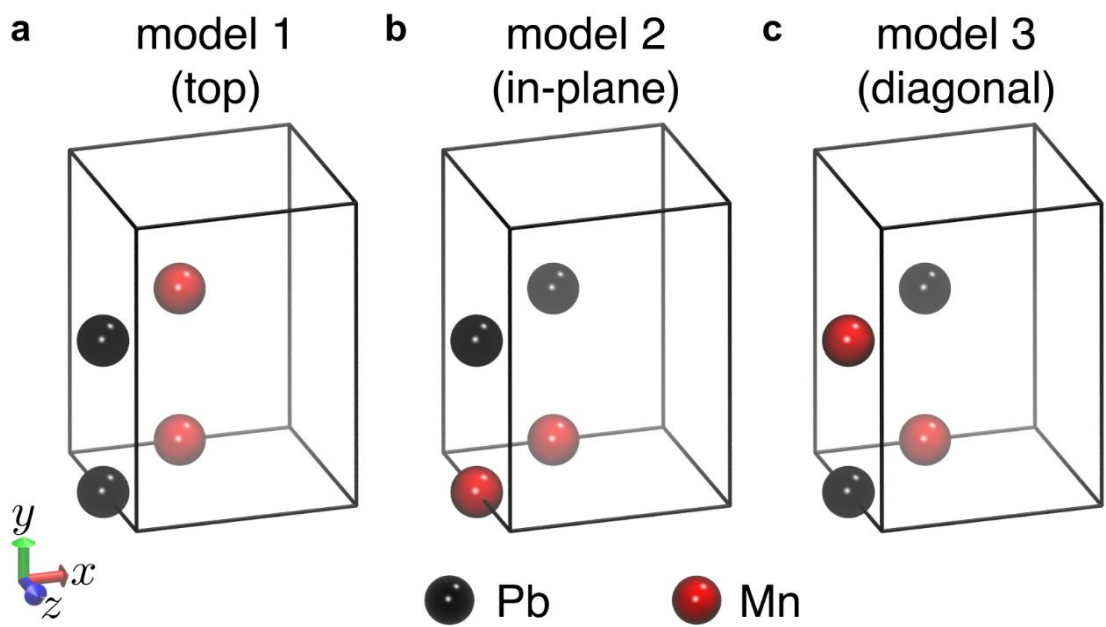

**Supplementary Figure 8 – Models of the Pb and Mn distributions in MAMn:PbI$_3$** Schematic drawings of three models of MAMn:PbI$_3$ containing pairs of Mn dopants in close proximity to each other in the 2×1×2 supercell studied by means of first-principles calculations. The three configurations investigated are referred to as top (**a**), in-plane (**b**), and diagonal (**c**). For clarity, only Pb (black) or Mn (red) atoms are shown and the unit cell of the undoped orthorhombic-phase MAPbI$_3$ is indicated by black lines.

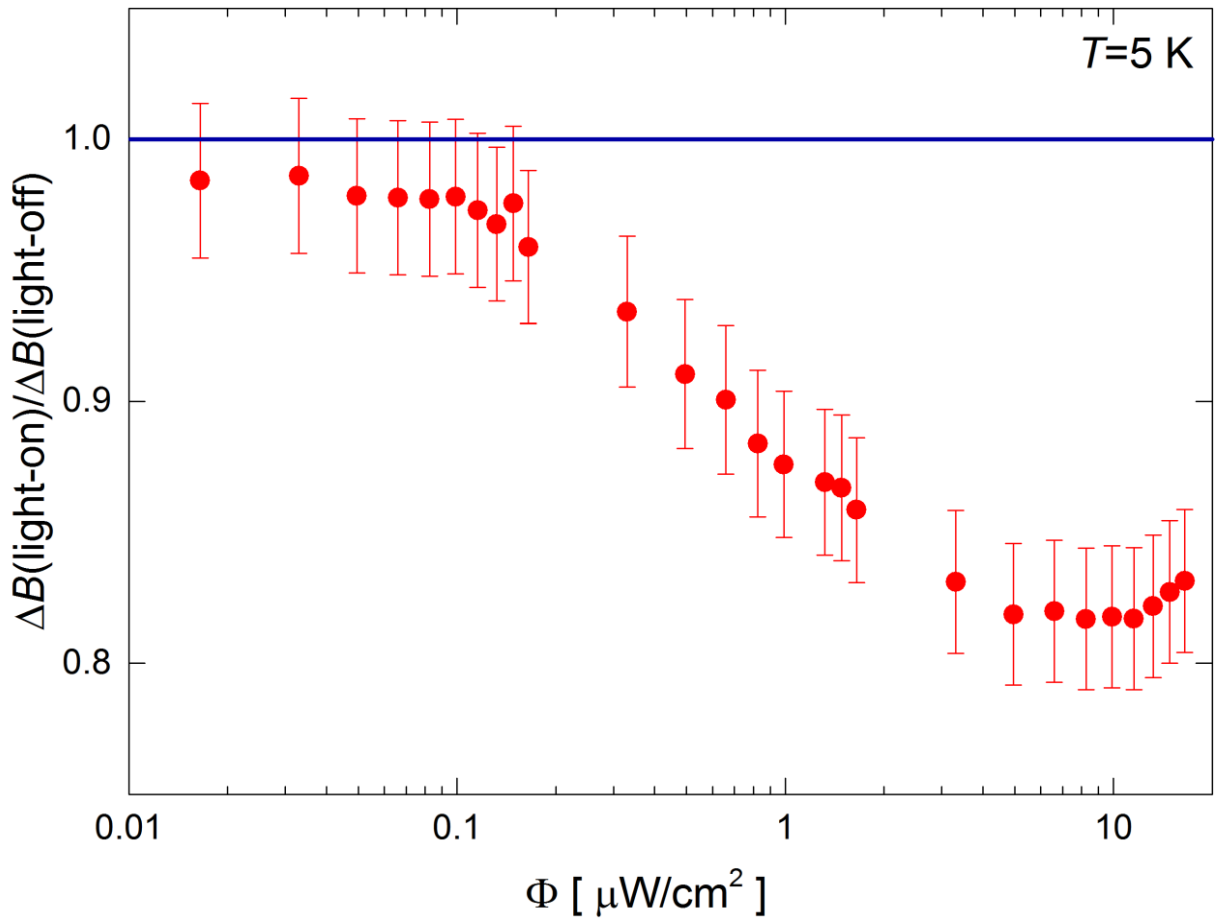

**Supplementary Figure 9 - Illumination intensity effect on MAMn:PbI$_3$ measured by ESR.** The change of the light-on ESR linewidth normalized to the linewidth in dark as the function of the illuminating red light intensity $\Phi$ at $T$=5 K. Above a threshold value, the FM part of the signal decreases monotonously in agreement with the intensity change seen in Fig. 3a in the main text.

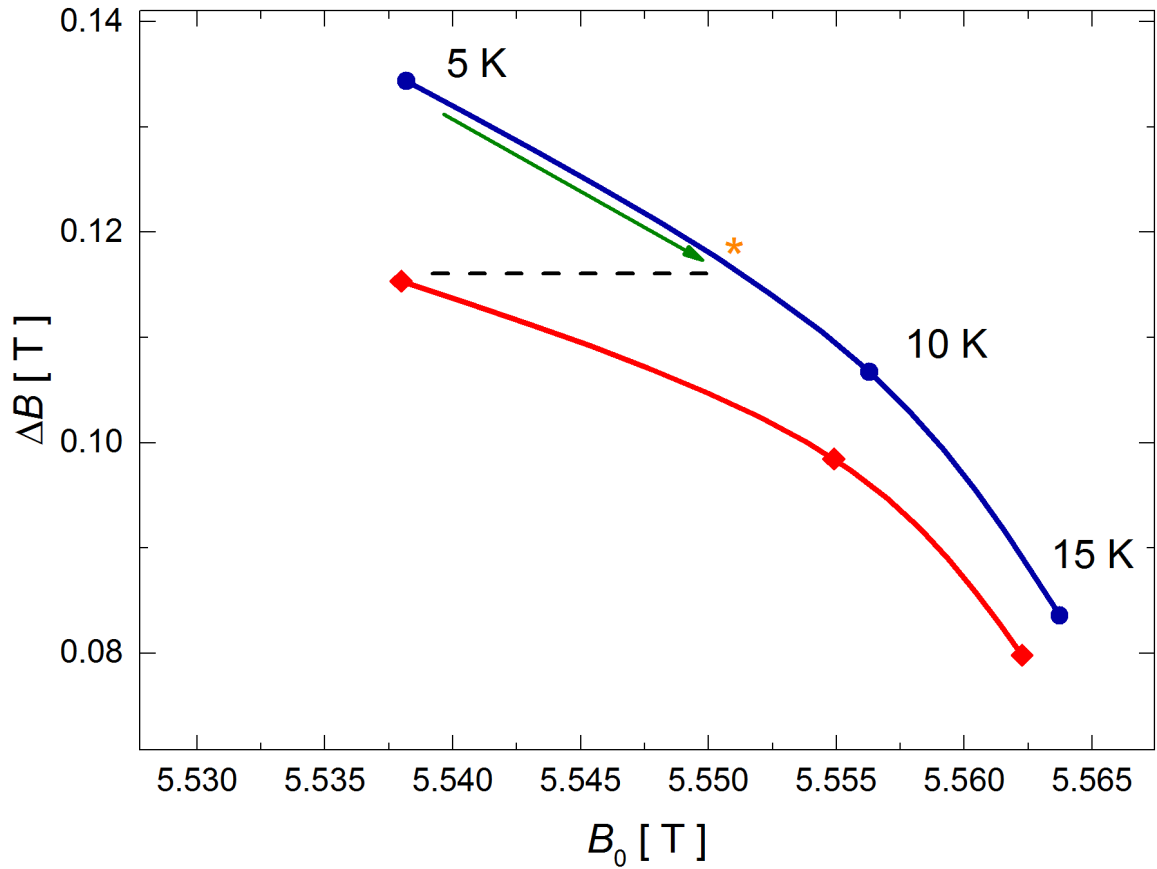

**Supplementary Figure 10- Magnetization melting of MAMn:PbI$_3$**: The resonance field $B_0$ increases with temperature, while the linewidth, $\Delta B$ of the MAMn:PbI$_3$ sample monotonically decreases (blue points are measured in dark, red points measured under 20 µW/cm$^2$ of light intensity at $T$=5, 10 and 15 K, the dashed line is a guide to the eye.) If the narrowing of $\Delta B$ was due to sample heating, one would move on the blue line in the direction indicated by the green arrow to the point shown by the orange star, and $B_0$ would move to a higher value.

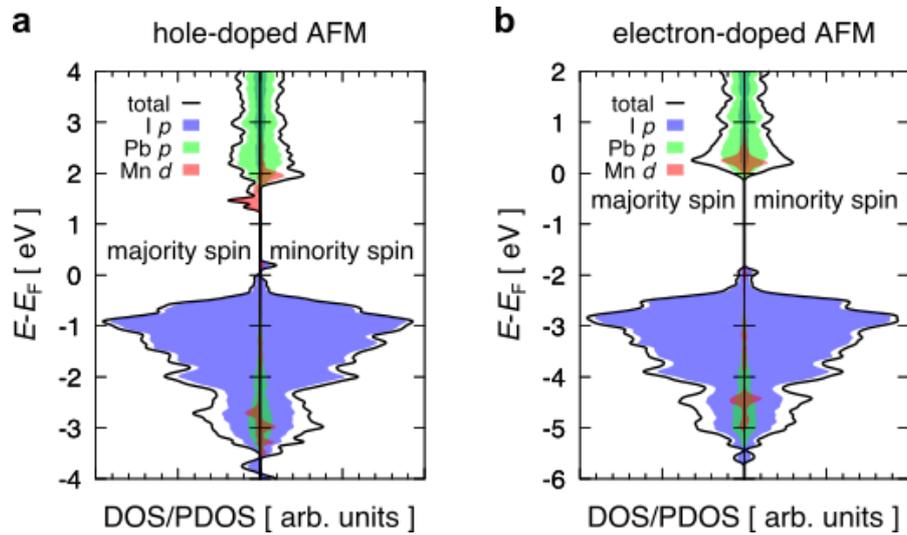

**Supplementary Figure 11 – Density of states plots for doped models of MAMn:PbI$_3$.** Total density of states (DOS) and projected density of states (PDOS) plots calculated from first-principles for the hole- (**a**) and electron-doped (**b**) in-plane model of MAMn:PbI$_3$ in the AFM ground state.

| | |
|---|---|
| Formula | $CH_3NH_3(Pb_{0.9}Mn_{0.1})$ |
| Cell settings | Tetragonal |
| Space group | *I4/mcm* |
| | |
| $a$ (Å) | 8.88078(18) |
| $c$ (Å) | 12.6981(3) |
| $\beta$ (degrees) | 90 |
| | |
| **Refinement** | |
| R, wR (observed) (%) | 2.43, 3.43 |
| R, wR (all) (%) | 2.53, 3.47 |
| $R\rho$, $wR\rho$ (%) | 3.11, 4.01 |
| $\Delta\rho_{max}$, $\Delta\rho_{min}$ (e Å$^{-3}$) | 0.86, -0.68 |

**Supplementary Table 1 - Structural characteristics of MAMn:PbI$_3$ at 293 K**